\documentclass[endfloats*,preprint,nobibnotes]{revtex4}

\usepackage{amsmath}
\usepackage{graphicx}
\usepackage{dcolumn}
\usepackage{bm}
\usepackage{color}

\newcolumntype{z}[1]{D{.}{.}{#1}}

\begin{document}




\title{Dissociation energy of the water dimer from Quantum Monte Carlo
calculations}

\author{I.~G.~Gurtubay$^{1,2}$ and R.~J.~Needs$^{1}$}

\affiliation{$^{1}$Theory of Condensed Matter Group, Cavendish
Laboratory, J.~J. Thomson Avenue, Cambridge CB3 0HE, United Kingdom}

\affiliation{$^{2}$Materia Kondentsatuaren Fisika Saila, Zientzia eta
Teknologia Fakultatea, Euskal Herriko Unibertsitatea, 644 Posta
kutxatila, E-48080 Bilbo, Basque Country, Spain }

\date{\today}

\begin{abstract}
We report a study of the electronic dissociation energy of the water
dimer using quantum Monte Carlo (QMC) techniques.  We have performed
variational quantum Monte Carlo (VMC) and diffusion quantum Monte
Carlo (DMC) calculations of the electronic ground state of the water
monomer and dimer using all-electron and pseudopotential approaches.
We have used Slater-Jastrow trial wave functions with B3LYP-like
single-particle orbitals, into which we have incorporated backflow
correlations.  When backflow correlations are introduced, the total
energy of the water monomer decreases by about 4-5 mHa, yielding a DMC
energy of $-$76.42830(5)~Ha, which is only 10~mHa above the
experimental value.  In our pseudopotential DMC calculations, we have
compared the total energies of the water monomer and dimer obtained
using the locality approximation with those from the variational
scheme recently proposed by Casula [Phys. Rev. B {\bf 74}, 161102(R)
(2006)].  The time step errors in the Casula scheme are larger and the
extrapolation of the energy to zero time step always lies above the
result obtained with the locality approximation. However, the errors
cancel when energy differences are taken, yielding electronic
dissociation energies within error bars of each other.  The
dissociation energies obtained in our various all-electron and
pseudopotential calculations range 
between 5.03(7) and 5.47(9)~kcal/mol
and are in
good agreement with experiment.  Our calculations give monomer dipole moments
which range between 1.897(2) and 1.909(4)~Debye and dimer
dipole moments which range between 2.628(6) and 2.672(5)~Debye.

\end{abstract}

\maketitle

\section{Introduction} \label{intro}

Water, as the main agent of all aqueous phenomena and an important
component of the vast majority of all chemical and biological
processes, has been the subject of many experimental and theoretical
studies.  The characteristic physical and chemical properties of water
stem from its strong polar hydrogen bonds.  In spite of the apparent
simplicity of hydrogen bonding -- a hydrogen atom bonded to an
electronegative group and a lone electron pair on another system --
understanding hydrogen bonding has proved very difficult.

One would hope to gain some insight by first investigating the water
monomer and then progressing to the water dimer and larger water
clusters, systematically including the many-body behaviour as the
number of molecules increases.  The study of the structure and
energetics of assemblies of water molecules poses severe challenges to
computational methods because they are held together by hydrogen bonds
with binding energies of only a few kcal/mol.

The hydrogen bond in the water dimer has been the subject of many
electronic structure studies
\cite{Feller-1992,Mas_Szalewicz-1996,Famulari_et_al-1998-2,
Halkier_et_al-1997,Klopper_et_al-2000,Park_et_al-2001,Xantheas_Burnham_Harrison-2002-2,
Tschumper_et_al-2002} since it represents the prototype of all
hydrogen-bonded systems.  In recent years the water dimer has been
studied using high-level quantum chemistry techniques, such as
second-order M{\o}ller--Plesset perturbation theory (MP2)
\cite{Xantheas_Burnham_Harrison-2002-2} and coupled-cluster CCSD(T)
methods \cite{Klopper_et_al-2000}.  These methods can treat electron
correlation effects quite accurately, but they are very expensive for
larger water clusters, as the costs of MP2 and CCSD(T) calculations
scale as $N^5$ and $N^7$, respectively, where $N$ is the number of
electrons.  The accuracy of the results depends significantly on the
quality of the basis set, and corrections are normally applied for the
effects of basis incompleteness.  Density-functional theory (DFT)
shows a more favourable scaling with system size, allowing the use of
larger basis sets resulting in smaller basis set errors.  However,
the calculated energies of the water monomer and dimer depend
significantly on the exchange-correlation functional used.

Quantum Monte Carlo (QMC) methods \cite{Foulkes_rmp-2001} represent an
alternative and promising way of treating electron correlation.  The
cost of calculating energies with QMC scales roughly as $N^3$,
allowing accurate calculations for large systems.  Moreover, QMC
algorithms are intrinsically parallel and Monte Carlo codes are easily
adapted to parallel computers.  As the availability of powerful
computers has increased, QMC has become a very attractive and
effective method for probing the electronic structure of molecules and
solids \cite{Drummond_diamondoids-2005,Maezono_etal-2007}.

In this paper we report variational and diffusion Monte Carlo (VMC and
DMC) calculations of the water monomer and dimer, using all-electron
(AE) and pseudopotential (PP) approaches.  We have used Slater-Jastrow
(SJ) and Slater-Jastrow-Backflow (SJB) wave functions, finding that
the introduction of backflow correlations retrieves a substantial
additional fraction of the correlation energy.  Our calculations yield
dissociation energies in very good agreement with experiment.  We
believe that our calculations form the most accurate QMC study of the
water monomer and dimer performed to date.

The rest of the paper is organized as follows.  In
Section~\ref{method} we give short reviews of the VMC and DMC methods,
the SJ and SJB wave functions, and the methods used for dealing with
non-local PPs within DMC.
In Section~\ref{details} we discuss the single-particle orbitals used
and some details of the calculations.  In Section~\ref{results} we
present and discuss our results.  The first part of
Section~\ref{results} describes results for the water monomer, the
second part gives results for the water dimer and its dissociation
energy into two water molecules, and we compare results from the two
different schemes used to calculate the non-local PP energy.  In the
third part, we report our results for the dipole moment of the monomer
and dimer.  Our conclusions are summarized in Section~\ref{concl}.

\section{VMC and DMC Methods}\label{method}

In VMC the ground-state energy is estimated as the expectation value
of the Hamiltonian with an approximate trial wave function, the
integrals being evaluated by importance-sampled Monte Carlo
integration.  The trial wave functions contain variable parameters,
whose values are obtained from an optimisation procedure formulated
within VMC.  There are no restrictions on the form of trial wave
functions that can be used, and VMC does not suffer from a fermion
sign problem.  However, the choice of the approximate trial wave
function is very important as it directly determines the accuracy and
statistical efficiency of the calculation.  Due to the difficulty of
preparing trial wave functions of equivalent accuracy for different
systems, the bias in the energy difference between systems is normally
significant.  We have used VMC methods mainly to optimize parameters
in the trial wave functions, and our most accurate calculations have
been performed with the DMC method.

In DMC the ground-state component of a trial wave function is
projected out by evolving an ensemble of electronic configurations
using the imaginary-time Schr\"odinger equation.  The fermionic
symmetry is maintained by the fixed-node (FN) approximation
\cite{Anderson-1976}, in which the nodal surface of the DMC wave
function is constrained to equal that of the trial wave function. The
FN DMC energy is higher than the exact ground-state energy, and
becomes equal to it when the fixed nodal surface is exact.  The
dependence of the DMC energy on the trial wave function is smaller
than in VMC, but in practice it is often significant.  It is therefore
desirable to be able to improve the nodal surfaces in order to reduce
the impact of the FN approximation.

The standard SJ wave function can be written as
\begin{equation}\label{eq:psiSJ}
\Psi_{\rm SJ}({\bf R}) = e^{J({\bf R})} \Psi_{\rm S}({\bf R}),
\end{equation}
where {\bf R} is a 3$N$-dimensional vector denoting the position ${\bf
r}_i$ of each electron.  The nodes of $\Psi_{\rm SJ}({\bf R})$ are
defined by the Slater part of the wave function, $\Psi_{\rm S}({\bf
R})$, which takes the form $\Psi_{\rm
S}\,=\,D_{\uparrow}\,D_{\downarrow}$, where $D_{\sigma}$ is a Slater
determinant of single particle orbitals of spin $\sigma$.

The Jastrow correlation factor, $ e^{J({\bf R})}$, contains
electron-electron, electron-nucleus and electron-electron-nucleus
terms, as described in Ref.~\onlinecite{Drummond_Towler_Needs-2004}.
The Jastrow factor keeps electrons away from each other and greatly
improves wave functions in general, but it does not modify the nodal
surface of the wave function.

One way of reducing the FN error is to alter the nodes of the wave
function by introducing backflow correlations
\cite{lopez-rios_etal-2006}, replacing the coordinates {\bf R} in the
Slater part of the wave function by the collective coordinates {\bf
X}, so that the SJB wave function reads
\begin{equation}\label{eq:psiiBF}
\Psi_{\rm SJB}({\bf R}) = e^{J({\bf R})} \Psi_{\rm S}({\bf X}).
\end{equation}
The new coordinates for each electron are given by
\begin{equation}\label{eq:psi}
{\bf x}_i = {\bf r}_i+\xi_i({\bf R}),
\end{equation}
$\xi_i$ being the backflow displacement of particle $i$, which depends
on the position of every electron in the system.  Details of the
specific form of the backflow function used can be found in
Ref.~\onlinecite{lopez-rios_etal-2006}.

\subsection{Pseudopotential DMC calculations}\label{PPs}

We use non-local (angular-momentum dependent) PPs.  Unfortunately
non-local potentials cannot be used directly in DMC calculations
because of the local character of the algorithm. Indeed, as we have
already mentioned, the starting point of the DMC algorithm is the
imaginary time Schr\"odinger equation. If the FN Hamiltonian, $H$,
contains a PP with a non-local part, $V_{\rm nl}$, and $E_T$ is a
trial energy, then the propagator for imaginary time $\tau$ is $<{\bf
R}| \rm{exp}[-\tau({H}-E_T)]|{\bf R'}>$. This propagator contains
terms of the form $<{\bf R}| \rm{exp}[-\tau V_{\rm nl}]|{\bf R'}>$,
which are not guaranteed to be positive for arbitrary ${\bf R}$ and
${\bf R'}$, and therefore cannot be interpreted as a probability
density.

One way of avoiding the possible change of sign is to use the PP
locality approximation (PLA) \cite{Mitas_Shirley_Ceperley-1991}.  In
this approximation the FN Hamiltonian is replaced by an effective
Hamiltonian in which the non-local part operates on the trial wave
function $\Psi_T$.  This technique is analogous to the FN
approximation. In this way the non-local part of the Hamiltonian is
replaced by a local term and the PLA effective Hamiltonian reads:
\begin{equation}\label{PLA-H}
{H}^{\rm PLA}_{{\bf R},{\bf R}}=K+V_{\rm loc}({\bf R})+ \frac{ \int
d{\bf R'}<{\bf R'} | V_{\rm nl} | {\bf R}> \Psi_T({\bf R'})}
{\Psi_T({\bf R})},
\end{equation} 
where $K$ is the kinetic energy operator and $V_{\rm loc}$ is the
local part of the PP.  In this approximation,
all of the matrix elements of 
the localized potential enter the branching factor of the DMC algorithm.  A
disadvantage of the PLA is that the energy may be higher or lower than
the exact value.

Casula has recently introduced a different scheme \cite{Casula-2006}
for treating non-local PPs within DMC.  In the Casula scheme (CS), the
FN Hamiltonian is substituted by an effective Hamiltonian
\begin{equation}\label{H-CS}
 \begin{cases}
{H}^{\rm CS}_{{\bf R},{\bf R}}= K+V_{\rm loc}({\bf R})+
				\sum\limits_{\bf R'}V_{{\bf R'} {\bf R}}^+
 & \\ 
{H}^{\rm CS}_{{\bf R'},{\bf R}}=<{\bf R'}| V_{\rm nl}|{\bf R}>
  & \text {if $V_{{\bf R'} {\bf R}}<0$ }\\
{H}^{\rm CS}_{{\bf R'},{\bf R}}=0 & \text {if $ V_{{\bf R'} {\bf R}}>0 $}\\ 
 \end{cases}
\end{equation}
where
\begin{equation}\label{V-CS}
V_{{\bf R'} {\bf R}}=\frac{\Psi_T({\bf R'})}{\Psi_T({\bf R})} <{\bf
R'} | V_{\rm nl}| {\bf R}>
\end{equation}
and 
\begin{equation}
V_{{\bf R'} {\bf R}}^+=\frac{1}{2}(V_{{\bf R'} {\bf R}}+|V_{{\bf R'} {\bf R}}|).
\end{equation}
From Eqs.~(\ref{H-CS}) and (\ref{V-CS}) we see that in the CS only the
positive matrix elements of the non-local potential are localized, and
they are absorbed into the branching factor.  On the other hand,
unlike in the PLA, the negative matrix elements of $V_{\rm nl}$
contribute to the drift-diffusion of the walkers.  In practice, the
standard DMC algorithm needs few changes: once a drift-diffusion move
is proposed and accepted or rejected, and after the walker has been
weighted, an extra displacement of the walker is performed according
to a transition probability that depends only on the negative matrix
elements of the non-local part of the PP.  This method has two
advantages over the PLA.  First, the ground state energy of the CS
Hamiltonian is an upper bound \cite{tenHaaf-1995} on the ground state
energy of the true FN Hamiltonian.  Second, whenever the negative
elements are large, the new displacement pushes the walkers away from
the attractive regions of the localized potential, which avoids
``population explosions'' in which walkers are trapped in the
attractive regions with large branching factors.  We found occasional
population explosions in our PLA calculations, which we dealt with by
restarting the calculations at an earlier move with a new random seed.
We found no such instabilities when using the CS.

\section{Details of the calculations}\label{details}

\begin{figure}[h!]
\includegraphics*[width=1.0\linewidth]{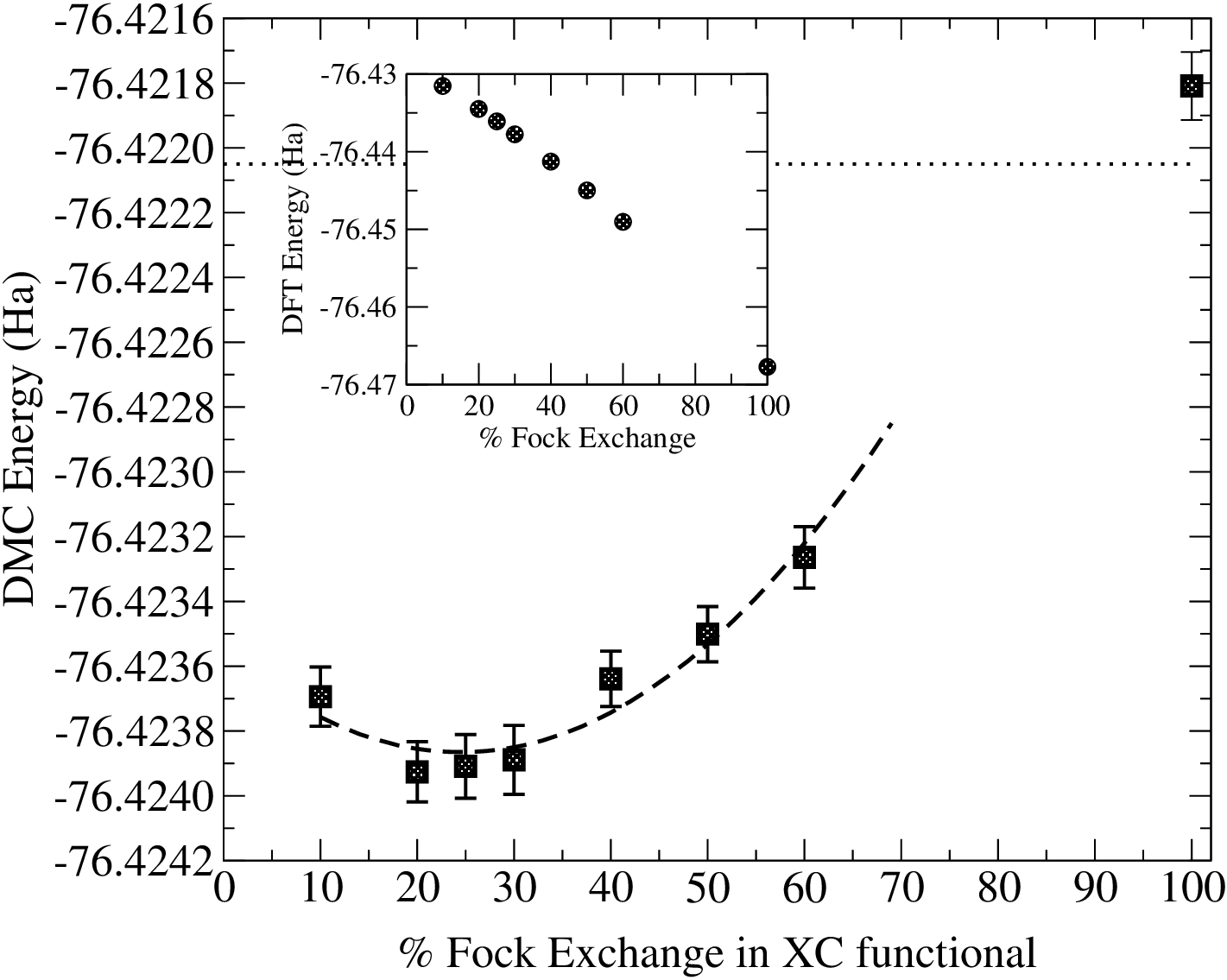}
\caption{\label{h2o_dmc_fit}
 Gurtubay et al., Journal of Chemical Physics}
\end{figure}

All the calculations have been performed using the \textsc{casino}
code \cite{casinomanual-2.0}.  The single-particle orbitals forming
the Slater determinants have been obtained from DFT calculations using
the \textsc{crystal98} code \cite{crystal98} with the Roos augmented
double zeta ANO (Roos aug-DZ-ANO) Gaussian basis set
\cite{Roos-ano-1990} used in its decontracted form.  Extensive tests
of the basis sets listed at
http://www.emsl.pnl.gov/forms/basisform.html showed that the Roos
basis set gives very good DMC energies.  The dependence of the DMC
energies on the orbitals is generally quite weak.  However, previous
calculations for neon, neon radicals \cite{Gurtubay_etal-2006} and
water \cite{Benedek_etal-2006} have shown that B3LYP orbitals give
slightly lower DMC energies than Hartree-Fock (HF) orbitals,
suggesting that the former can improve the nodal surface of the wave
function.  Based on this, we have investigated whether one can obtain
further improvements in the DMC energies by optimising the form of the
exchange-correlation (XC) functional used to generate the
single-particle orbitals.  Becke's three parameter B3LYP XC functional
\cite{Becke-1993} may be written as \cite{crystal98}:
\begin{equation}\label{Exc}
E_{\rm xc} = (1-A) (E_x^{\rm LDA}+B\, E_{\rm x}^{\rm Becke}) +
            A\,E_{\rm x}^{\rm HF} + (1-C)\,E_{\rm c}^{\rm VWN} +
            C\,E_c^{\rm LYP},
\end{equation}
where $A$ determines the amount of Fock exchange, and $B$ and $C$
determine the amounts of non-local exchange and correlation,
respectively.  In Eq.~(\ref{Exc}), the Vosko-Wilk-Nusair correlation
potential (VWN) is used to extract the local part of the LYP
correlation potential.  The usual B3LYP functional corresponds to
$A=0.2$, $B=0.9$ and $C=0.81$.  In principle we could try and minimize
the DMC energy with respect to $A$, $B$ and $C$, but this would be
very expensive.  Setting $A=1$ in Eq.~(\ref{Exc}), we have obtained a
DMC energy for the water molecule of $-$76.4218(1)~Ha
\footnote{The value in brackets denotes the standard error in the mean
in the last figure.  In this case, for example, $-$76.4218(1) means
$-$76.4218$\pm$0.0001.}, which is very close to the value of
$-$76.42205(8)~Ha obtained with HF orbitals.  This suggested that the
contribution from the correlation part (terms involving $C$) does not
significantly affect the DMC energy.
We therefore decided to search for the amount of Fock exchange ($A$)
which minimizes the DMC energy, while keeping $B$ and $C$ equal to
their standard B3LYP values.  The results are shown in
Fig.~\ref{h2o_dmc_fit}.
Note that the DMC energy (squares) is much less sensitive to the
changes in the XC functional than the DFT energy (circles).  The
dashed line represents the best quadratic fit to the DMC energies for
$A<$0.6.  The lowest DMC energy is obtained when 25\% Fock exchange is
used in the above functional.  We therefore used $A=0.25$ to generate
the single-particle orbitals for the Slater parts of all of the trial
wave functions used in this work.  In what follows, we refer to these
as B3LYP-like orbitals.

In our AE calculations, the single-particle molecular orbitals close
to the nuclei have been corrected using the scheme described in
Ref.~\onlinecite{Ma_cusp-2005} so that each orbital obeys the
appropriate cusp condition \cite{Kato-1957} at each nucleus, ensuring
that the local energy is finite at each nucleus.  In our PP
calculations, we have used the two schemes described in
Section~\ref{PPs} with PPs generated within Hartree-Fock theory,
including scalar relativistic effects
\cite{Trail_Needs-2005a,Trail_Needs-2005b}. These PPs have been found
to work very well in conjuction with QMC methods
\cite{Gurtubay_etal-2006}.  One of the advantages of using PPs is that
they avoid the short-range variations in the wave function near the
nuclei, and hence we can use larger time steps than is possible in AE
calculations.  We have performed calculations for different time steps
and then extrapolated to the limit of zero time step by fitting to
quadratic functions.  Population-control biases have been checked by
running with different population sizes, and no such bias was
detected.

The parameters in the Jastrow and backflow functions were obtained by
minimizing the variance of the local energy
\cite{Kent_Needs_Rajagopal-1999,Drummond_Needs-2005}, except for the
AE calculations with SJB wave functions.  In this case, the Jastrow
and backflow parameters were obtained by minimizing the mean absolute
deviation of the set of the local energies from the median.

\section{Results and discussion}\label{results}


The calculations for the water monomer were carried out in the
experimental equilibrium geometry with r$_{\rm OH}$ = r$_{\rm OH'}$ =
0.9572~${\rm \AA}$ and $\angle = 104.52^{\rm o}$.  This geometry was
used in previously reported QMC calculations
\cite{Luchow_Anderson_Feller-1997,Benedek_etal-2006}.

\begin{figure}[htb]
\includegraphics*[width=0.70\linewidth]{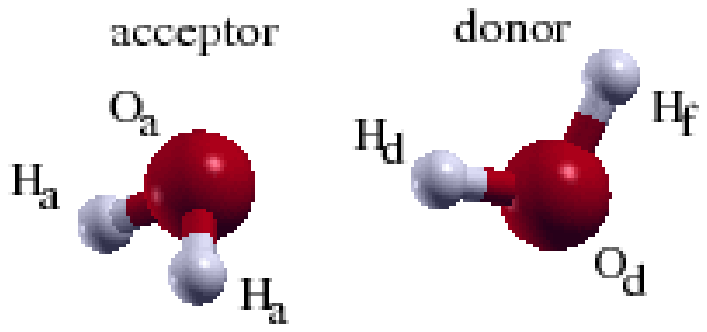}
\caption{\label{dimergeometry} 
 Gurtubay et al, Journal of Chemical Physics}
\end{figure}

For the water dimer we have chosen the geometry derived by Klopper
{\it et al.} \cite{Klopper_et_al-2000} from CCSD(T) calculations,
extrapolating to the basis set limit using MP2-R12 results. The
optimized water dimer geometry (see Fig.~\ref{dimergeometry}) yields a
final equilibrium oxygen-oxygen distance of 2.912~${\rm\AA}$.  When
the dimer is formed, some structural deformation occurs in each of the
monomers: a slight change in the angle between the inter-oxygen vector
and the plane of the proton-accepting monomer occurs and the bond
angle for the proton acceptor widens by about 0.5$^{\rm o}$.  We have
neglected these small changes, assuming that their effect on the
intermolecular binding energy is small.  The electronic dissociation
energy in the equilibrium geometry, $D_e$, with respect to
dissociation into two isolated monomers, is defined as
\begin{equation}\label{De}
D_e=D_o+{\rm ZPE (dimer)}-2\,{\rm ZPE(monomer)},
\end{equation}
where $D_o$ is the experimental dissociation energy and ZPE denotes
the zero point energy of the nuclei.

\subsection{Water monomer}

Our AE SJ and SJB QMC results with B3LYP-like orbitals for the water
monomer are given in Table~\ref{AEmonomer}, together with energies
calculated using other {\it ab initio} methods.
The expansion order of the polynomials and the spin dependencies 
used in the Jastrow factor as described in 
Ref.~\onlinecite{Drummond_Towler_Needs-2004} were the following:
N$_u$=N$_{\chi}$=6, N$_f^{eN,ee}$=3, S$_u$=1 and S$_{\chi}$=S$_f$=0.
This led to 128 parameters in the Jastrow factor.
 When backflow correlations were incorporated the total
number of parameters to be optimized increased to 338, 
corresponding to the following expansion order of the polynomials in
the backflow term (Ref.~\onlinecite{lopez-rios_etal-2006}):
N$_{\eta}$=N$_{\mu}$=6, N$_{\phi}^{eN,ee}$=3, S$_{\eta}$=S$_{\phi}$=1
and S$_{\mu}$=0.
 We have
mentioned in Section~\ref{method} that the VMC energy depends strongly
on the trial wave function.  This is illustrated by
Table~\ref{AEmonomer}: when backflow correlations are included, the
VMC energy for the water molecule [$-$76.4034(2)~Ha] decreases by
nearly 26~mHa.
Our DMC energy extrapolated to zero time step with the SJ wave function
(DMC-SJ), $-$76.42376(5)~Ha, is nearly 1~mHa lower than that recently
reported by Benedek {\it et al.} \cite{Benedek_etal-2006} using true
B3LYP orbitals and lies about 14~mHa above the experimental result of
$-$76.438~Ha.  When backflow correlations are introduced, the DMC
energy decreases by a further 4.5~mHa, which is 25\% of the
correlation energy missing with the SJ wave function, yielding a DMC
energy of $-$76.42830(5)~Ha.  Note that, as mentioned in
Section~\ref{method}, the difference between the energies obtained
with the SJ and SJB trial wave functions is smaller at the DMC level
than at the VMC level.

L\"{u}chow and Fink \cite{Luchow_Fink-2000} performed DMC calculations
using a pair natural orbital configuration-interaction (PNO-CI) trial
wave function consisting of 300 determinants and a Jastrow factor,
obtaining an energy of $-$76.429(1)~Ha, which is within error bars of
our single-determinant SJB DMC result.  For completeness,
Table~\ref{AEmonomer} also gives the DMC energy obtained by Casula
{\it et al.} \cite{Casula_Attacalite_Sorella-2004} using a trial wave
function consisting of a product of an antisymmetrized geminal power
(AGP) wave function and a Jastrow factor, which is nearly 7~mHa above
our best DMC result.  Table~\ref{AEmonomer} also gives the DMC
energies calculated with SJ wave functions and HF nodes from
Refs.~\onlinecite{Gurtubay_etal-2006} and
\onlinecite{Benedek_etal-2006}.  As was pointed out in
Section~\ref{details}, both of these energies are slightly higher than
the DMC results obtained using B3LYP and B3LYP-like orbitals, which
suggests that the hybrid DFT functional yields better nodes than the
HF orbitals.  The small difference between the two DMC-HF values given
in Table~\ref{AEmonomer} arises from the use of different basis sets.
For comparison, we have also included energies from a CCSD(T)-R12
calculation \cite{Muller_Kutzelnigg_Noga-1997} and a recently reported
Correlation Energy Extrapolation by Intrinsic Scaling (CEEIS)
calculation \cite{Bytautas_Ruedenberg-2006} which, although not
variational, give results very close to the ``exact'' one.

\begin{table}[ht!]
\begin{tabular}{l z{3} z{12} l   z{3} c  }
\hline\hline
\multicolumn{1}{l}{Method}&\multicolumn{1}{c}{}&
\multicolumn{1}{c}{Total energy (Ha)} &\multicolumn{1}{c}{Basis set}&
\multicolumn{1}{c}{}&\multicolumn{1}{c}{Ref.} \\
 \hline
DFT         & &-76.4361& Roos Aug-DZ-ANO&& This work  \\
VMC-SJ      & & -76.3773(2)&Roos Aug-DZ-ANO && This work \\
VMC-SJB     & & -76.4034(2)&Roos Aug-DZ-ANO&& This work \\
DMC-SJ      & & -76.42376(5)&Roos Aug-DZ-ANO&& This work \\
DMC-SJB     & & -76.42830(5)&Roos Aug-DZ-ANO&& This work \\
\hline
DMC-HF      & & -76.42102(4)& 6-311++G(2d,2p)&& \onlinecite{Gurtubay_etal-2006}\\
DMC-HF      & & -76.4219(1)&Roos Aug-DZ-ANO&& \onlinecite{Benedek_etal-2006} \\
DMC-B3LYP   & &-76.4230(1)&Roos Aug-DZ-ANO&& \onlinecite{Benedek_etal-2006} \\
DMC-PNO-CI  & &-76.429(1)&STO           && \onlinecite{Luchow_Fink-2000} \\
DMC-AGP     & &-76.4175(4)&STO-DZ         &&\onlinecite{Casula_Attacalite_Sorella-2004}  \\
CCSD(T)-R12 & & -76.4373&Dunning cc-pV6Z&& \onlinecite{Muller_Kutzelnigg_Noga-1997} \\ 
CEEIS       & & -76.4390(4)&CBS limit      &&\onlinecite{Bytautas_Ruedenberg-2006}\\
Exact       & & -76.438&&&  \onlinecite{Feller_Boyl_Davidson-1987}   \\
\hline\hline
\end{tabular}
\parbox{\linewidth}{
\caption{\label{AEmonomer} Total energy of the water monomer
calculated with different {\it ab initio} methods, and the exact
non-relativistic static point nuclei value.
CBS limit stands for extrapolation to the complete basis set limit.
}}
\end{table}

\subsection{Water dimer and electronic dissociation energy}

As illustrated in Fig.~\ref{dimergeometry}, the two hydrogen atoms of
the acceptor monomer (H$_{\rm a}$) are not equivalent to the donor
hydrogen (H$_{\rm d}$) and free hydrogen (H$_{\rm f}$) atoms in the
donor subunit, and the two oxygen atoms (O$_{\rm a}$ and O$_{\rm d}$)
are inequivalent.  However, we have obtained slightly lower DMC
energies by using Jastrow and backflow functions which treat all four
hydrogen atoms and both oxygen atoms as equivalent.  Treating atoms as
equivalent in this way reduces the number of variable parameters, and
presumably this results in the better performance of the stochastic
optimization procedure.  When the H and O atoms are treated as
equivalent, the SJ trial wave functions for the dimer contains 128
parameters (for both the AE and PP calculations) while the SJB
calculations have been performed with 390 parameters for the AE
calculations and 286 parameters for the PP calculations.
These number of parameters correspond to
 N$_u$=N$_{\chi}$=6, N$_f^{eN,ee}$=3, S$_u$=S$_f$=1 and S$_{\chi}$=0 
in both the AE and PP Jastrow factors. In the backflow term,
N$_{\eta}$=N$_{\mu}$=6, S$_{\eta}$=S$_{\phi}$=1, S$_{\mu}$=0
in both AE and PP calculations, and
N$_{\phi}^{eN,ee}$=3 and N$_{\phi}^{eN,ee}$=2 in the AE and PP cases,
respectively.

\begin{figure}[h!]
\includegraphics*[width=0.95\linewidth]{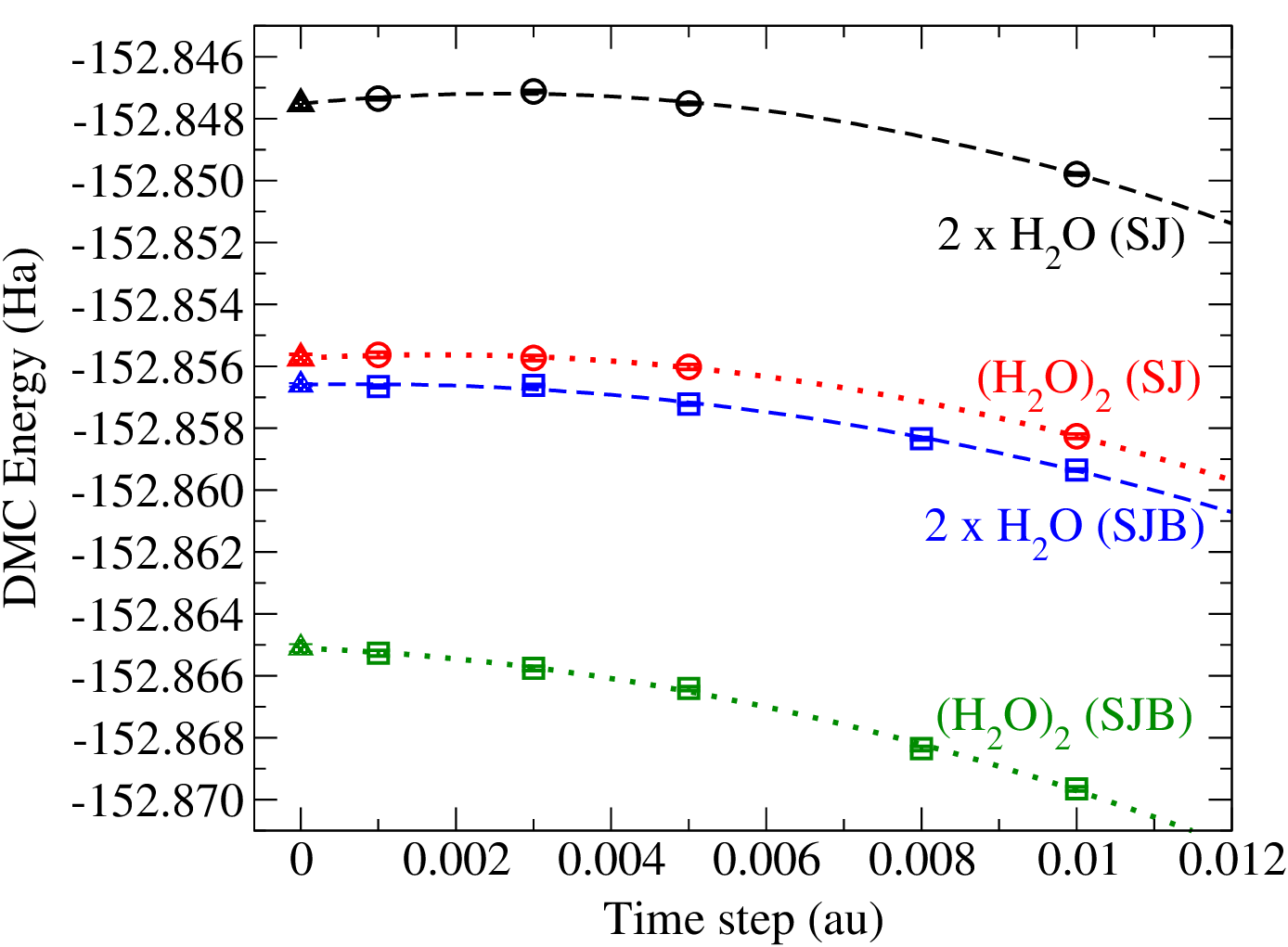}
\caption{\label{totalE-AE}
 Gurtubay et al, Journal of Chemical Physics}
\end{figure}

Figure~\ref{totalE-AE} shows the AE DMC energies of two water monomers
and the water dimer as a function of time step.
The zero-time-step extrapolations of the fitted curves are shown by
filled triangles of the same color as the original symbols. The DMC-SJ
and DMC-SJB values shown in Table~\ref{AEmonomer} correspond to the
black and blue triangles in this figure, {\it i.e.}, the extrapolated
values from the curves labelled 2$\times$H$_2$O(SJ) and
2$\times$H$_2$O(SJB), respectively.  As we mentioned previously for
the water monomer, the DMC energy decreases by about 4.5~mHa when
backflow correlations are introduced.  Figure~\ref{totalE-AE} shows
that the reductions in energy from including backflow correlations are
essentially independent of the time step for both the monomer and
dimer.

\begin{figure}[h!]
\includegraphics*[width=0.95\linewidth]{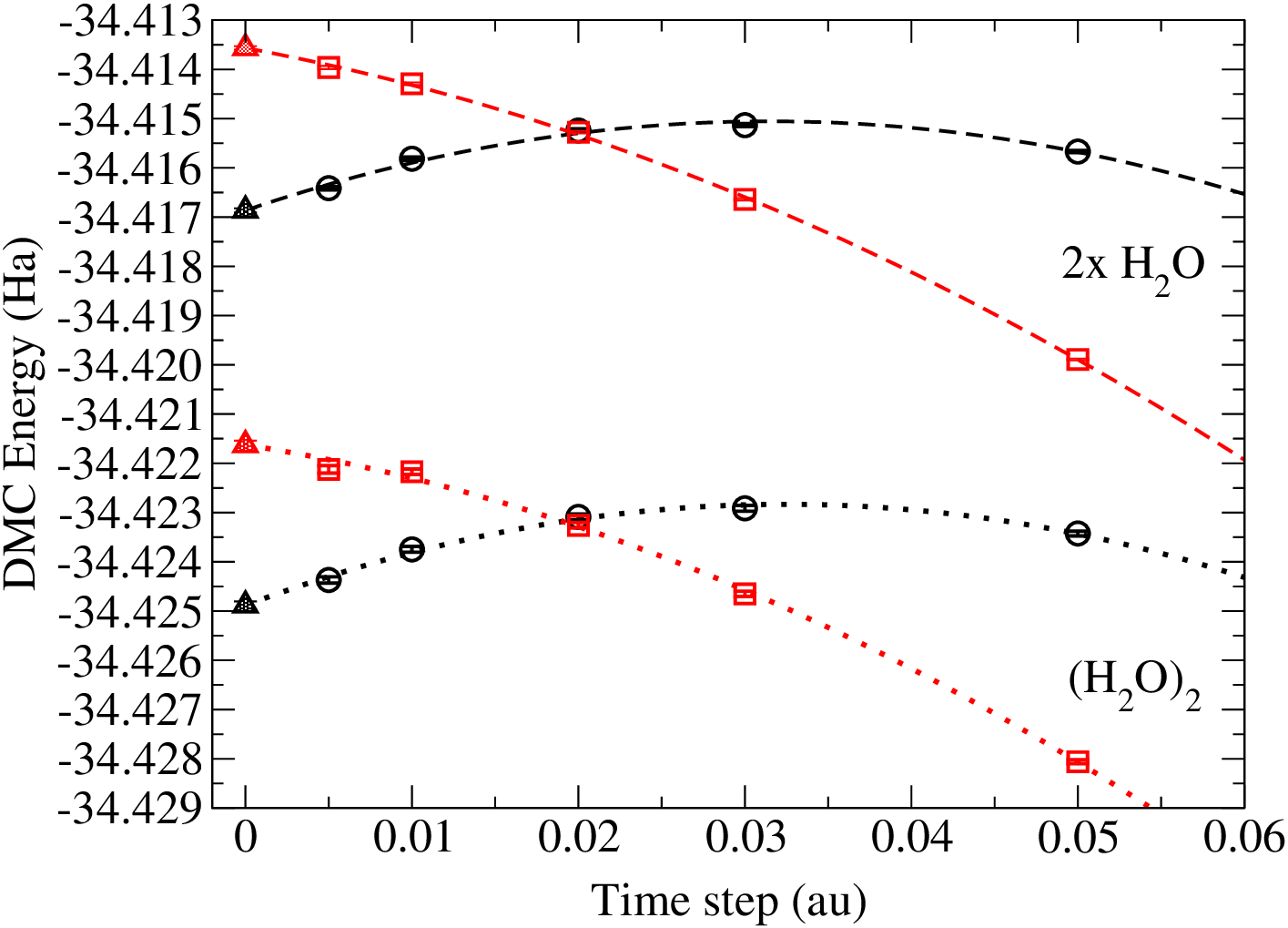}
\caption{\label{casula-comp}
 Gurtubay et al, Journal of Chemical Physics}
\end{figure}

Figure~\ref{casula-comp} shows PP SJ DMC energies for two water
monomers and the dimer as a function of the time step. 
The CS and PLA curves cross at a time step of about $0.02$ au.  The
extrapolated SJ DMC energy of two water monomers obtained within the
CS scheme is less than 4~mHa above the PLA value.  It is not
surprising that the CS energies are above the PLA ones, because the CS
scheme gives an upper bound on the true energy, while the PLA scheme
does not.  The energy of the water monomer as a function of the time
step in both the PLA and CS schemes can be found in
Table~\ref{compdata}.

\begin{table}[h!]
\begin{tabular}{z{10}  z{10}   z{10} }
\hline\hline
\multicolumn{1}{c}{}&\multicolumn{2}{c}{Energy (Ha)}\\
\cline{2-3}
\multicolumn{1}{c}{Time step (au)}&\multicolumn{1}{c}{PLA}&\multicolumn{1}{c}{CS} \\
\hline
0.000  & -17.20843(4) & -17.20678(4) \\
0.005  & -17.20821(3) & -17.20698(3) \\ 
0.010  & -17.20791(3) & -17.20714(3) \\
0.020  & -17.20762(3) & -17.20763(3) \\
0.030  & -17.20757(3) & -17.20832(2) \\
0.050  & -17.20783(2) & -17.20994(2) \\
\hline\hline
\end{tabular}
\parbox{\linewidth}{
\caption{\label{compdata} Comparison of PP SJ DMC energies obtained
within the PLA and CS schemes.}}
\end{table}

\begin{figure}[htb]
\includegraphics*[width=0.95\linewidth]{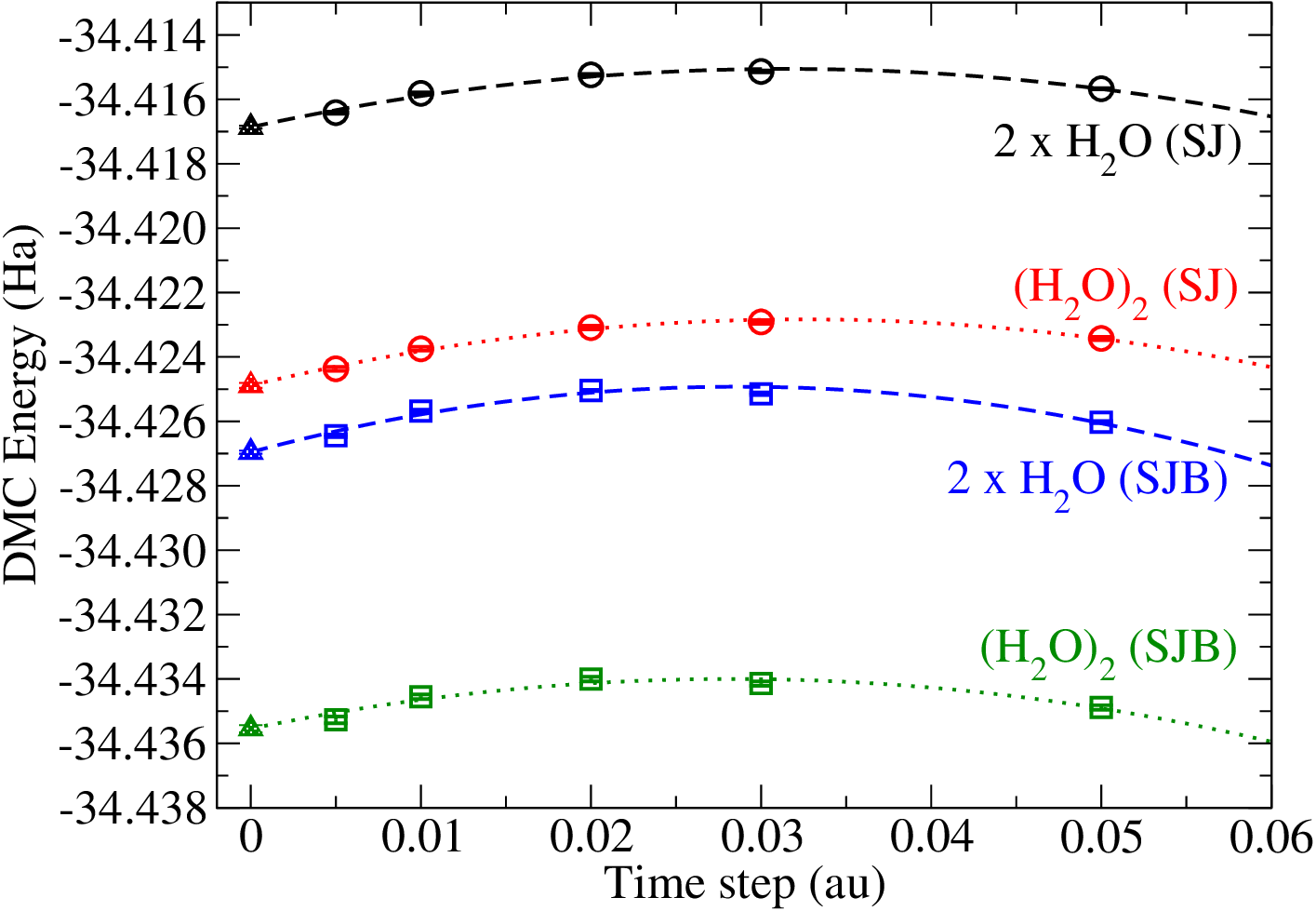}
\caption{\label{totalE-PP}
 Gurtubay et al, Journal of Chemical Physics}
\end{figure}

\begin{figure}[htb]
\includegraphics*[width=0.95\linewidth]{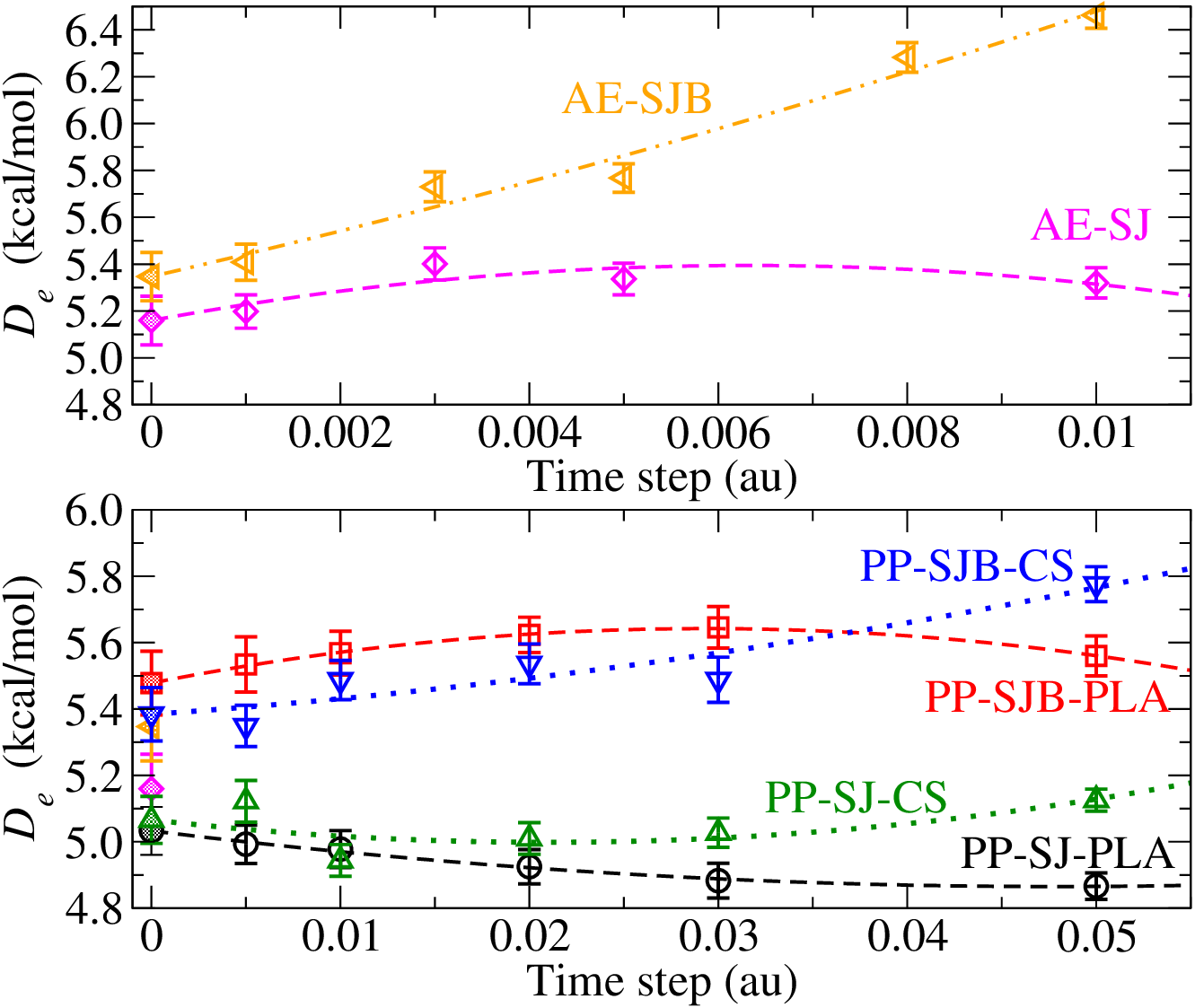}
\caption{\label{dissoc}
 Gurtubay et al, Journal of Chemical Physics}
\end{figure}

In Fig.~\ref{totalE-PP} we have summarized the total energies of the
PP water monomer and dimer energies with SJ and SJB wave functions
within the PLA. The CS yields a similar figure (not shown) with curves
parallel to the red squares of Fig.~\ref{casula-comp}.  As in
Fig.~\ref{totalE-AE}, including backflow correlation in the monomer
and dimer results in an essentially rigid shift of the corresponding
SJ curve to lower energies.  Furthermore, because the black and red
curve corresponding to the SJ calculations, and the blue and green
curves, corresponding to the SJB calculations, have nearly the same
shape, the time step errors largely cancel when computing the $D_e$ of
the water dimer.  This cancellation is illustrated in
Fig.~\ref{dissoc}, where we have plotted $D_e$ as a function of the
time step for the four combinations of SJ/SJB wave functions within
the AE/PP approaches from Figs.~\ref{totalE-AE} and \ref{totalE-PP}.

The top panel of Fig.~\ref{dissoc} shows the AE values of $D_e$. The
magenta diamonds (orange triangles) show the results with SJ (SJB)
wave functions as a function of time step.
We fitted the energies to quadratic functions and extrapolated to zero
time step, obtaining the values of $D_e$ given in
Table~\ref{De-table}.

\begin{table}[h!]
\begin{tabular}{l z{3}   z{10} c}
\hline\hline
\multicolumn{1}{c}{Method}&\multicolumn{1}{c}{}&\multicolumn{1}{c}{ $D_e$ (kcal/mol)} &\multicolumn{1}{c}{Ref.} \\
\hline
AE-SJ               & & 5.16(10) & This work\\
AE-SJB              & & 5.35(10) & This work\\
PP-SJ (PLA)         & & 5.03(7)  & This work\\
PP-SJ (CS)          & & 5.07(7)  & This work\\
PP-SJB (PLA)        & & 5.47(9)  & This work\\
PP-SJB (CS)         & & 5.38(8)  & This work\\
\hline
AE-HF-SJ            & & 5.02(18) & \onlinecite{Benedek_etal-2006}\\
AE-B3LYP-SJ         & & 5.21(18) & \onlinecite{Benedek_etal-2006}\\
CCSD(T)	            & & 5.02(5)  & \onlinecite{Klopper_et_al-2000}\\
Experiment + Theory & & 5.44(70) & \onlinecite{Curtiss_et_al-1979}\\
Experiment + Theory & & 5.00(70) & \onlinecite{Mas_etal-2000}\\
Experiment + Theory & & 5.14     & \onlinecite{Leforestier_et_al-2002}\\
\hline\hline
\end{tabular}
\parbox{\linewidth}{
\caption{\label{De-table} Top: DMC dissociation energies (kcal/mol)
obtained in this work with SJ and SJB wave functions using B3LYP-like
orbitals within the AE and PP (PLA and CS) approaches.  Bottom:
Results from other {\it ab initio} calculations and from 
``Experiment + Theory''. No error bar was reported
in Ref.~\onlinecite{Leforestier_et_al-2002}.}}
\end{table}

The high degree of cancellation of the time step errors shown in
Fig.~\ref{dissoc} is remarkable.  Our AE-SJ data show smaller time
step errors than those reported in Ref.~\onlinecite{Benedek_etal-2006}
with HF single-particle orbitals.  When backflow correlations are
introduced, the cancellation of time step errors at larger time steps
is not as good.  This can also be observed in Fig.~\ref{totalE-AE}
where the green line bends more than the blue one at larger time
steps.  Nevertheless, the extrapolated AE values of $D_e$ for the SJ
wave function [5.16(10)~kcal/mol] and the SJB wave function
[5.35(10)~kcal/mol] are within error bars of each other.  For the PP
calculations we have seen that the PLA and CS schemes give
significantly different time step errors.  However, the time step
errors are very similar for the monomer and dimer (see
Fig.~\ref{casula-comp}). Therefore the values of $D_e$ computed within
the PLA and CS schemes are very similar at each time step.  We find
$D_e =$ 5.03(7)~kcal/mol [5.07(7)~kcal/mol] for SJ wave functions
within the PLA [CS], and $D_e =$ 5.47(9)~kcal/mol [5.38(8)~kcal/mol]
when backflow correlations are introduced.  We have also found that,
in both our AE and PP calculations, backflow correlations tend to
increase the value of $D_e$ slightly.

Our DMC results range between 5.03(7) and 5.47(9)~kcal/mol and are
similar to the DMC values of $D_e$ reported in
Ref.~\onlinecite{Benedek_etal-2006} (with HF and true B3LYP orbitals)
and the CCSD(T) result of Klopper {\it et al.}
\cite{Klopper_et_al-2000} of 5.02(5)~kcal/mol.

For many years the experimental value of $D_e$ has been taken to be
5.44$\pm$0.7~kcal/mol. This value was derived from measurements of the
enthalpy of dimerization of water corrected with a theoretical
estimate of the ZPE \cite{Curtiss_et_al-1979}.  Mas {\it
  et al.} \cite{Mas_etal-2000} have recently calculated a more accurate value
of the ZPE and have estimated $D_e$ to be 5.00$\pm$0.7~kcal/mol.
Leforestier {\it et al.} \cite{Leforestier_et_al-2002} have reported
a value of $D_e$ of 5.14~kcal/mol, correcting for the ZPE by
determining  potential energy surfaces for the monomer and dimer via direct
inversion of spectroscopic data. 

\subsection{QMC dipole moment of the water monomer and dimer}

We have calculated the mean dipole moment, $\mu$, of the water monomer
and dimer.  The dipole moment operator does not commute with the
Hamiltonian and therefore the error in the DMC mixed estimate
($\mu_{\rm{DMC}}$) is linear in the error in the wave function.  This
error can be reduced to a quadratic error by using the so called
extrapolated estimator, $2\mu_{\rm{DMC}}-\mu_{\rm{VMC}}$, where
$\mu_{\rm{VMC}}$ is the VMC estimate of the dipole moment. Both VMC
and DMC estimates must, of course, be calculated using the same trial
wave function.

\begin{figure}[h!]
\includegraphics*[width=0.95\linewidth]{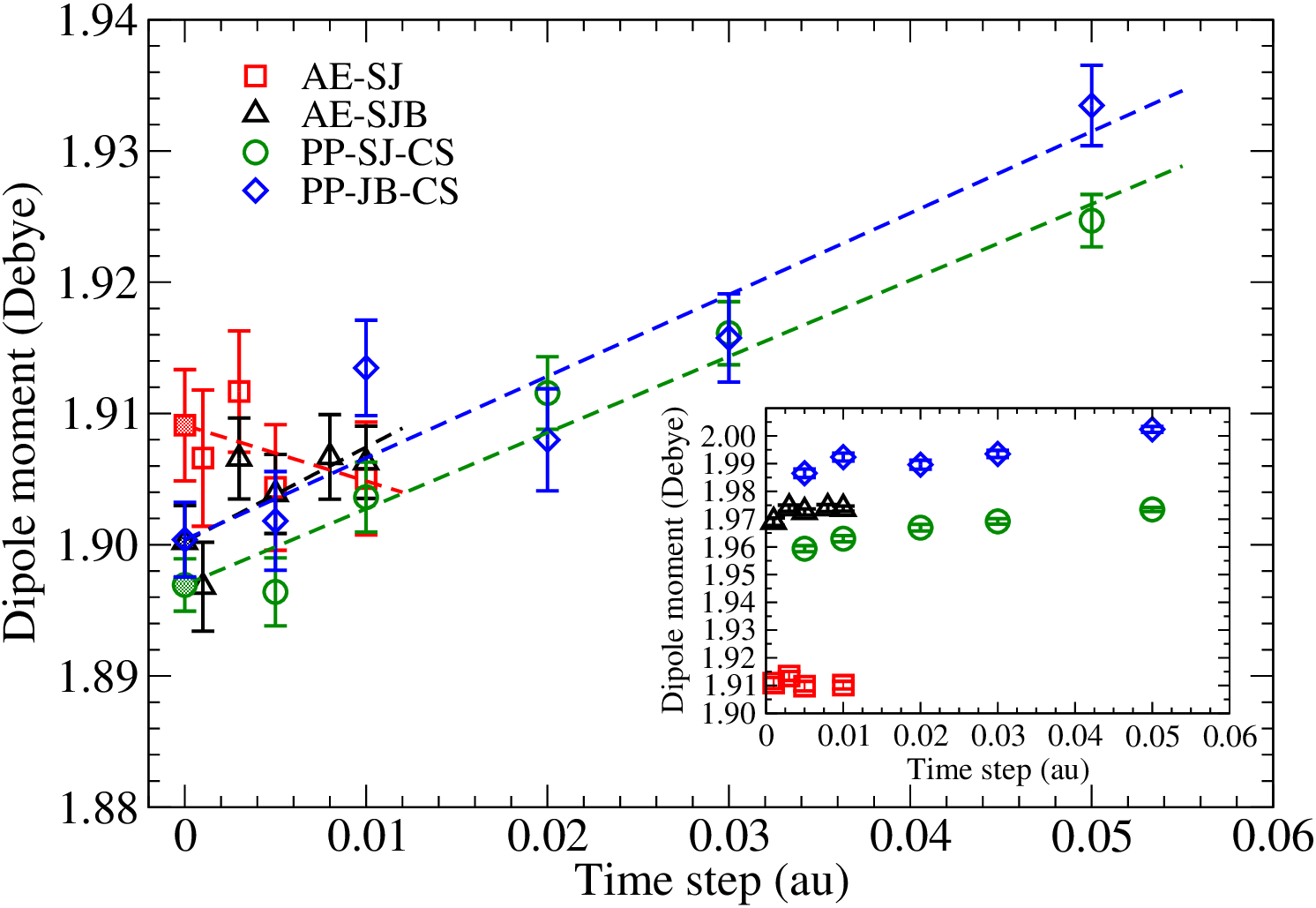}
\caption{\label{dip-mono}
 Gurtubay et al, Journal of Chemical Physics}
\end{figure}

Figure~\ref{dip-mono} shows our extrapolated estimator results for the
dipole moment of a water molecule as a function of time step.  Red
squares (black triangles) show AE SJ (SJB) results while green circles
(blue diamonds) represent PP SJ (SJB) results calculated with the CS.
Dashed lines of the same color as the original symbols are linear fits
used to find the zero-time-step dipole moment. Our results are
summarized in Table~\ref{table-dip-mono} together with the results of
other {\it ab initio} calculations.  Our results for the monomer are
slightly above (0.05~Debye) the experimental value, as are the other
theoretical values.  For comparison, the inset of Fig.~\ref{dip-mono}
shows the DMC mixed estimates of the dipole moment for each of the
four cases investigated. As we have mentioned above, the DMC estimate
has a larger dependency on the trial wave function and this can be
observed in the larger differences between the data sets displayed in
the inset. However, when the extrapolated estimator is considered the
error is reduced and the results come into closer agreement.

\begin{table}[h!]
\begin{tabular}{l    z{10} c}
\hline\hline
\multicolumn{1}{c}{Method} & \multicolumn{1}{c}{$\mu$ (Debye)} & \multicolumn{1}{c}{Ref.}\\
\hline
AE-SJ                   & 1.909(4) & This work\\
AE-SJB                  & 1.900(3) & This work\\
PP-SJ (CS)              & 1.897(2) & This work\\
PP-SJB (CS)             & 1.900(3) & This work\\
\hline
aug-cc-VDZ MP2          & 1.868    & \onlinecite{Gregory_etal-2006}\\
aug-cc-pVQZ MP2 and MP4 & 1.87     & \onlinecite{Xantheas_dunning-1993}\\
aug-cc-pVDZ MP2-R12     & 1.883    & \onlinecite{Klopper_et_al-1995}\\
Experiment              & 1.855    & \onlinecite{Gregory_etal-2006}\\
\hline\hline
\end{tabular}
\parbox{\linewidth}{
\caption{\label{table-dip-mono} Top: Dipole moment of the water
monomer, evaluated using the extrapolated estimator, and also
extrapolated to zero time step.  Bottom: Other {\it ab initio} results
and the experimental value.}}
\end{table}

We have arrived at similar conclusions for the dipole moment of the
water dimer (figure not shown).  The extrapolated estimator values of
the dipole moment of the water dimer, extrapolated to zero time step,
are summarized in Table~\ref{table-dip-dimer}.  Our AE results are
lower than the experimental value \cite{Dyke_et_al-1977}, 
but within 0.02~Debye of it, and our PP
results are higher than the experimental value, but within 0.03~Debye of
it.

\begin{table}[h!]
\begin{tabular}{l    z{10} c}
\hline\hline
\multicolumn{1}{c}{Method} & \multicolumn{1}{c}{$\mu$ (Debye)} & \multicolumn{1}{c}{Ref.}\\ 
\hline
AE-SJ             & 2.628(6) & This work\\
AE-SJB            & 2.635(7) & This work\\
PP-SJ (CS)        & 2.670(5) & This work\\
PP-SJB (CS)       & 2.672(5) & This work\\
\hline
aug-cc-VDZ MP2    & 2.683    & \onlinecite{Gregory_etal-2006}\\
aug-cc-pVDZ B3LYP & 2.635    & \onlinecite{Xantheas-1995}\\
Experiment        & 2.643    & \onlinecite{Dyke_et_al-1977}\\
\hline\hline
\end{tabular}
\parbox{\linewidth}{
\caption{\label{table-dip-dimer} As in Table~\ref{table-dip-mono}, but
for the water dimer.}}
\end{table}

\section{Conclusions}\label{concl}

We have performed all-electron and pseudopotential QMC calculations of
the total energy of the water monomer and dimer using SJ and SJB wave
functions with B3LYP-like single-particle orbitals.  The introduction
of backflow correlations recovers a significant amount of correlation
energy and we have obtained a final DMC energy (extrapolated to zero
time step) for the water molecule of $-$76.42830(5)~Ha.
We have performed a careful study of the time step errors and have
shown that they largely cancel when energy differences, such as the
electronic dissociation energy of the water dimer, are considered.  We
have compared two different schemes for treating non-local PPs within
DMC: the pseudopotential localization approximation and a scheme
recently proposed by Casula.  Although the total energies from the two
schemes differ significantly, these differences are essentially the
same in the monomer and dimer, and therefore they cancel when
evaluating $D_e$.  Our extrapolated AE and PP results for the $D_e$
show good agreement with each other when both SJ and SJB wave
functions are used, and they are all within the error bars of the
values determined from experiment.  The total spread of our AE and PP
values of $D_e$ is only 0.44 kcal/mol.  We have also reported QMC
estimates of the dipole moments of the water monomer and dimer.  Our
AE and PP extrapolated DMC results yield a dipole moment of the
monomer within 0.05~Debye of the experimental value, while for the dimer
the AE results are within 0.02~Debye of experiment and the PP results are
within 0.03~Debye of experiment.

\acknowledgments I.G.G. gratefully acknowledges financial support from
the Basque Government through grant BFI04/183, the EC 6$^{\rm th}$
framework Network of Excellence NANOQUANTA (NMP4-CT-2004-500198), and
the Ministerio de Educaci\'on y Ciencia (FIS2006-01343 and
CSD2006-53).  R.J.N. gratefully acknowledges support from the
Engineering and Physical Sciences Research Council (EPSRC) of the
United Kingdom.  Computational facilities were provided by the
Cambridge High Performance Computing Service and the SGI/IZO-SGIker
UPV/EHU (supported by the National Program for the Promotion of Human
Resources within the National Plan of Scientific Research, Development
and Innovation-Fondo Social Europeo, MCyT and the Basque Government).

\bibliography{labur,qmc,water,quantum_chem,gral}
\cleardoublepage

{\bf Figure captions}
\\
\vspace*{1cm}

Fig. 1:
 DMC energy (squares) as a function of the
percentage of Fock exchange in the exchange-correlation B3LYP-like
functional [$A$ in Eq.~(\ref{Exc})] used to obtain the single-particle
orbitals for the SJ wave functions.  The same Jastrow factor of 128
parameters was used in all the calculations.  The dashed line shows a
quadratic fit to the points around the minimum.  The dotted line
represents the DMC energy with HF orbitals.  Inset: DFT energy as a
function of the percentage of Fock exchange.
\\
\vspace*{2cm}

Fig. 2: (Color online)
The equilibrium
structure of the water dimer from Klopper {\it et al.}
\cite{Klopper_et_al-2000} The ``a'' stands for acceptor, ``d'' for
donor and ``f'' for free.
\\
\vspace*{2cm}

Fig. 3: (Color online)
 All-electron total energies
of two water monomers and the water dimer as a function of time step,
calculated with SJ (circles) and SJB (squares) wave functions.  Dashed
and dotted lines show quadratic fits to the monomer and dimer
energies, respectively.  The filled triangles represent the
extrapolation to zero time step.  The statistical error bars are of
order $8\times 10^{-5}$~Ha.
\\
\vspace*{2cm}

Fig. 4: (Color online)
SJ PP energies for two
water monomers and for the water dimer, calculated with the PLA (black
circles) and the CS (red squares). Dashed (dotted) lines are fits for
the monomer (dimer). Triangles denote the energies extrapolated to
zero time step.
\\
\vspace*{2cm}

Fig. 5: (Color online)
 As in Fig.~\ref{totalE-AE}, but using PPs within the PLA.
\\
\vspace*{2cm}

Fig. 6: (Color online)
 Dissociation energy of the
water dimer as a function of time step.  The curves represent time
step extrapolations and filled symbols give the zero time step $D_e$.
Top panel: AE calculations.
Bottom panel: PP calculations.
For comparison, the AE values extrapolated to zero time step are also
shown (with the same colors and symbols as in the top panel).
\\
\vspace*{2cm}

Fig. 7: (Color online)
Extrapolated estimates of the
dipole moment of the water molecule as a function of the time
step. Filled symbols correspond to zero time step values obtained from
linear fits (dashed lines). Inset: DMC mixed estimator.

\end{document}